\documentclass[prc,aps,preprint,showpacs,nofootinbib,showkeys,eqsecnum]{revtex4}            
\usepackage{epsfig}
\newcommand{\noi}{\noindent}

\def\be{\begin{equation}}
\def\ee{\end{equation}}
\def\bea{\begin{eqnarray}}
\def\eea{\end{eqnarray}}
\def\Journal#1#2#3#4{{#1} {\bf #2}, #3 (#4)}

\def\PRC{{ Phys. Rev.} \bf C}

\def\AJ{ Astrophys. J.\,\,}

\def\JPBAMOP{J. Phys. B: At. Mol. Opt. Phys.\,\,}

\begin{document}
%
%
\title{
{\large{\bf Reply to the Comment on "Influence of protons on the
capture of electrons by the nuclei of $^{7}Be$ in the Sun"}}}
\author{V. B. Belyaev}
\affiliation{Bogolyubov Laboratory of Theoretical Physics, Joint
Institute for Nuclear Research, Dubna  141980, Russia }
\author{M. Tater}
\affiliation{Institute of Nuclear Physics ASCR, CZ--250 68 \v{R}e\v{z}, Czech Republic }
\author{E.~Truhl\'{\i}k}
\affiliation{Institute of Nuclear Physics ASCR, CZ--250 68 \v{R}e\v{z}, Czech Republic }

\begin{abstract}
We show that the arguments against our paper raised by B. Davids
{\it et al.} are either irrelevant or incorrect.
\end{abstract}

\noi 

\noi \hskip 1.9cm 

\maketitle

\section{Introduction}

In our reply to the Comment \cite{dgj} we first summarize shortly
results of our work \cite{btt} and then we shall analyze the
critical remarks of the Comment. Let us note here that we were
interested in \cite{btt} only in the mechanism of the elementary
process of the capture of the electron from the continuum \be
^{7}Be+e^- +p\rightarrow ^{7}Li+\nu_e+ p\,,\label{TR}\ee and did not
discuss the possible plasma effects. In the binary reaction \be
^{7}Be+e^- \rightarrow ^{7}Li+\nu_e\,,\label{BR1}\ee these effects
were found small \cite{b2,b7,b14}. Since the mean distance $R_0$
between protons is about $3\times 10^4$ fm and the Debye radius
$R_D$ is about $4\times 10^4$ fm, one can expect that these effects
will be small also for the reaction (\ref{TR}). Therefore we shall
ignore all the discussion of the Comment concerning the description
of the plasma.

Necessity to consider the reaction (\ref{TR}) follows from the fact
that in the standard theory of the $pp$  cycle the destruction of
the nucleus $^{7}Be$ takes place both in the binary reactions
(\ref{BR1}) and \be
p\,+\,^{7}Be\,\rightarrow\,^{8}B\,+\,\gamma\,.\label{BR2}  \ee

Let us stress that we study \cite{btt} the reaction (\ref{TR})
quantum-mechanically using the solution of the Schroedinger equation
applied in \cite{b3} for the description of the system of three
charged particles in the continuum. It was shown \cite{b3} for the
case of two heavy and one light particle that the three-body wave
function can be expanded in a small parameter $\epsilon$. In
principle, the heavy particles are allowed to interact also
strongly. In our case
$\epsilon\,\approx\,(m_e/m_p)^{(1/2)}\,\approx\,0.0233$. Here $m_e$
($m_p$) is the electron (proton) mass. The first term of this
expansion is \mbox{$\Psi_0(\vec r, \vec R)\,=\, \Psi^C_1(\vec
R)\Psi^C(\vec r, Z=Z_1+Z_2)$}, where $\Psi^C_1(\vec R)$ is the
Coulomb wave function describing the relative motion of the proton
and $^{7}$Be and satisfies the equation, \be -\frac{{\not
h}^2}{2m_r}\Delta_{\vec R}\,\Psi^C_1(\vec R)+\frac{Z_1
Z_2e^2}{R}\Psi^C_1(\vec R)\,=\,E_1\Psi^C_1(\vec R)\,.\label{SE1}\ee
Here $m_r$ is the reduced mass of the proton-$^{7}$Be system. In its
turn, $\Psi^C(\vec r, Z=Z_1 +Z_2)$ is the Coulomb wave function of
the electron moving in the continuum in an effective Coulomb
potential of the charge $Z=Z_1 + Z_2$ and satisfies the equation,
\be -\frac{{\not h}^2}{2m_e}\Delta_{\vec r}\,\Psi^C(\vec
r)-\frac{Ze^2}{r}\Psi^C(\vec r)\,=\,E\Psi^C(\vec r)\,.\label{SE2}\ee
So in contrast to the binary reaction (\ref{BR1}), in the three-body
initial state (\ref{TR}) the motion of the electron occurs in the
Coulomb field with an effective charge $Z=5$ and the vector $\vec r$
points to the position of the electron relative to the center of
mass of the proton-$^{7}$Be system. It means that to calculate the
electron capture by $^{7}$Be in the ternary reaction (\ref{TR}), one
needs to know its wave function $\Psi^C(\vec r, Z=5)$ for the value
of $|\vec r|$ equal to the distance between the center of mass of
the system proton-$^{7}$Be and the position of $^{7}$Be which
corresponds to zero distance between the electron and $^{7}$Be. This
is the main qualitative result of \cite{btt} which also contains the
quantitative comparison of the effect of the electron density in the
ternary state (\ref{TR}), given by the function $\Psi^C(\vec r,
Z=5)$, with the electron density in the binary reaction (\ref{BR1}).
According to Table I \cite{btt}, the ratio of these effects is in
the Sun of the order of 10 \%, but it can be of the order 1 in the
dense stars, as can be seen from our Fig.\,2 \cite{btt}.

\section{Analysis}

Let us now analyze the arguments of the Comment. In our opinion, the
content of the Comment contains the three groups of contradictory
arguments. To the first group one can relate arguments of the type
"three-body mechanism of the electron capture by $^{7}$Be is, in
fact, a binary one". In the second group of the arguments, the
existence of the reaction (\ref{TR}) is accepted de bene esse in a
sense that if even the three-body mechanism works, both the wave
functions and the calculations are in our work totally wrong. To the
end of the Comment, Davids {\it et al.} solve the problem by an
argument that the effect of the reaction (\ref{TR}) has already been
taken into account in \cite{b14} by describing the binary reaction
(\ref{BR1}) within the framework of the formalism of the equilibrium
plasma and using the Monte Carlo technique to include the
interaction of the static protons with the electron and the $^{7}$Be
nucleus. Non-biased reader can easily follow the evolution of the
"proofs" of Davids {\it et al.} from the full negative of the
existence of the three-body effects to the full acceptance of their
presence in the equilibrium plasma. We show below that these
arguments are either contradictory or have no relation to our work.

Being not able to refute the fact of the absence of works dealing
with the explicit three-body mechanism of the capture of the
electrons in the continuum by $^{7}$Be, Davids {\it et al.} try to
prove that the reaction (\ref{TR}) is the binary process. If one
omits completely the fundamental difference in the kinematics
between the ternary- and binary collisions and also excludes the
influence of the Coulomb interaction of the proton with the electron
and $^{7}$Be, then the ternary process converts into the binary one.
However, these assumptions, admitted by Davids {\it et al.}, have no
relation to the real situation. Following the logics which is
difficult to follow, Davids {\it et al.} compare the mean distance
between the particles in the plasma with the range of the weak
interaction. Evidently, this comparison has no sense and has no
relation to the role of the three-body mechanism of the reaction,
because the structure of the weak Hamiltonian has no influence on
the formation of the initial state.

The essence of the point of view advocated by Davids {\it et al.}
can be understood from the text presented at the end of the third
paragraph at p.\,1: {\it "In fact, the only influence such a proton
can have on the electron capture rate is electromagnetic, by
affecting the density of electrons at the $^{7}$Be nucleus.
Therefore it is incorrect to think of this as a ternary reaction.
Rather it is a binary reaction in a plasma environment."} This
philosophy was implemented in the calculations \cite{b14} where the
influence of the static protons surrounding $^{7}$Be on the capture
rate of the electrons in the binary reaction (\ref{BR1}) was taken
into account. It is clear that the static protons do not change the
binary feature of the reaction and can cause only the change of the
capture rate. But our point of view is that the reaction (\ref{BR1})
is not the only possible mode of the electron capture and the
channel (\ref{TR}) also occurs. In this case, the proton possesses
the explicit dynamical degree of freedom which means that the
Hamiltonian describing the initial state contains not only the
Coulomb interaction between the proton, electron and $^{7}$Be
nucleus, but in addition to the kinetic energy terms of the electron
and $^{7}$Be also such a term for the proton is present. This
changes the situation essentially because instead of only one Jacobi
coordinate for the binary electron-$^{7}$Be system one should
introduce two Jacobi coordinates characterizing the three-body
proton-electron-$^{7}$Be system. Then instead of the two-body
Schroedinger equation one is to solve the three-body one which was
done in \cite{b3}.

As it follows from the analysis \cite{b3} discussed above, the
three-body process (\ref{TR}) due to the presence of the proton in
the vicinity of $^{7}$Be, possessing the explicit dynamical degree
of freedom, results in the capture of the electron by an effective
charge $Z$=5 instead of $Z$=4. Moreover, the presence of such a
proton in the final state causes that, in contrast to the binary
reaction (\ref{BR1}), the resulting neutrino spectrum is not
monoenergetic but continuous one. This is a typical feature of the
neutrino spectrum in a reaction resulting in a many-body final
state, like the triton beta-decay. It is clear that these features
of the ternary reaction (\ref{TR}) cannot be simulated by the
screening corrections as calculated by the Monte Carlo simulations
in Ref.\,\cite{b14} for the binary reaction (\ref{BR1}), as wished
by Davids {\it et al.} in the last paragraph of the Comments.
Moreover, the claim {\it "Equilibrium statistical mechanics takes
care of the three-body and other effects."} is characteristic, as
all the text of the Comments, by the interchange of notions
consisting in that we speak about the elementary reaction of the
capture whereas Davids {\it et al.} talk about the description of
the plasma effects. However they do not recognize that the argument
about the presence of the three-body effects in the equilibrium
plasma used against considering independently the three-body
elementary process, can be applied also to the binary mechanism of
the capture. In other words, following the logics of Davids {\it et
al.}, the binary elementary processes should be also excluded from
the treatment which does not make any sense.

Discussing the quality of our wave functions Davids {\it et al.}
argue in the last paragraph on p.\,3: {\it "Its equation 2.3 is a
poor approximation to the three body wave function 2.2 in the limit
of interest, namely when the electron and the $^{7}$Be nucleus are
spatially coincident and the proton  is some 30 000 fm away from the
other two particles. Clearly, this approximation grows worse and
worse as the proton-$^{7}Be$ separation $R$ increases and the
magnitude of the Coulomb wave function describing the relative
motion of the proton and $^{7}Be$ vanishes."} Probably, here Davids
{\it et al.} speak about some other system and have in mind the wave
function different from the one given by Eq.\,(2.3), because the
asymptotic of this wave function coincides exactly with the
asymptotic of the Coulomb wave function that describes the proton
motion in the Coulomb field with the charge $Z$=3, as it should be
in the configuration in which the electron occurs in the vicinity of
the $^{7}$Be nucleus \cite{b3}. Moreover this qualitative reasoning
of Davids {\it et al.} is not supported by any quantitative results.
On the contrary, as it is seen from Eq.\,(2.12) \cite{btt}, the
integration over the variable $R$  excludes the contribution of the
wave functions from large and short distances due to the presence of
the exponential damping factor. Consequently, possible deformation
of the proton-nucleus wave functions due to the screening cannot
influence essentially the results.

Let us note here that our wave functions describe correctly not only
the $p-e-^{7}Be$ system, but also the well known $p-e-p$ reaction,
providing for the electron function the Fermi function with the
effective charge $Z$=2 and for the $pp$ system the standard quantum
mechanical wave function obtained by solving the Schroedinger
equation with both the strong and Coulomb interactions included.
Namely these wave functions were used to describe the $p-e-p$
reaction by Bahcall and May \cite{bama} forty years ago.

In the following text {\it "The paper asserts that the Coulomb wave
function of the electron in the field of the combined charges of the
proton and $^{7}Be$, $\Psi^C(\vec r,Z=Z_1+Z_2)$, defines the
probability of $^{7}Be$ electron capture. In fact, this is the
Coulomb wave function describing the relative motion of an electron
and $^{8}B$, and is only applicable when the proton is closer to the
$^{7}Be$ than the electron is. In electron capture this
approximation breaks down since the electron-$^{7}Be$ separation
must vanish in order for the capture to occur."}, Davids {\it et
al.} show clearly that they missed completely the essence of our
work and try to palm something off on us which is complete nonsense
and what has nothing to do with what we have done. Let us stress
once more that the wave function $\Psi^C(\vec r, Z=Z_1+Z_2)$ does
describe the motion of the electron in the effective field of the
proton-$^{7}Be$ system with $Z=Z_1+Z_2=5$, and the vector $\vec r$
does point from the center of mass of the proton-$^{7}$Be system to
the electron. It follows that the value of the electron coordinate
$|\vec r|=R/7$ is the distance of $^{7}$Be to the center of mass of
the p-$^{7}$Be system. As to the act of the electron capture by
$^{7}$Be, our picture is : The modulus squared
$|\Psi^C(r=R/7,Z=5)|^2$ provides the probability that the electron
can be registered at the point $|\vec r|=R/7$ where $^{7}$Be is
situated and can capture the electron. Since the expansion (2.2) and
Eq.\,(2.3) \cite{btt} are valid in the whole space of the variables
$\vec r$, $\vec R$, we conclude here that the above quoted reasoning
of Davids {\it et al.} is just a bleak fiction.

In summary, we have shown that the arguments  by Davids {\it et al.}
that the reaction (\ref{TR}) can take place in the solar plasma only
with the proton as a spectator the influence of which has already
been taken into account in \cite{b14} by calculating the screening
to the binary reaction (\ref{BR1}) are false. On the other hand we
are aware that our model calculations \cite{btt} cannot be
considered as a substitute for full calculation of this process that
only can provide the reliable information on the size of its rate.

\end{document}